\documentclass[11pt]{article}
\usepackage{moriond,epsfig}

\def\beq{\begin{equation}} 
\def\eeq{\end{equation}} 
\def\bea{\begin{eqnarray}} 
\def\eea{\end{eqnarray}}
\def\bq{\begin{quote}} 
\def\eq{\end{quote}}

\def\gappeq{\mathrel{\rlap {\raise.5ex\hbox{$>$}} {\lower.5ex\hbox{$\sim$}}}}

\def\lappeq{\mathrel{\rlap{\raise.5ex\hbox{$<$}} {\lower.5ex\hbox{$\sim$}}}}

\def\Journal#1#2#3#4{{#1} {\bf #2}, #3 (#4)}

\def\PLB{{\em Phys. Lett.}  B}

\def\PRD{{\em Phys. Rev.} D}

\def\mco{\multicolumn}

\def\ra{\rightarrow}

\def\ko{K^0}

\def\be{\begin{equation}}
\def\ee{\end{equation}}
\def\bea{\begin{eqnarray}}
\def\eea{\end{eqnarray}}

\begin{document}
\vspace*{4cm}
\title{A SUMMARY FROM THE THEORIST'S POINT OF VIEW}

\author{G. Altarelli}

\address{Theoretical Physics Division,CERN, CH-1211 Geneva Switzerland}
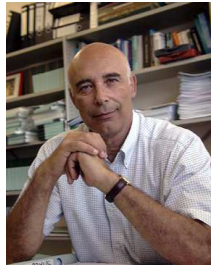
\maketitle\abstracts{
1. Introduction\\
2. Astrophysics and Cosmology \\
3. Neutrino Oscillations\\
4. Higgs and New Physics Searches\\
5. Flavour Physics and CP Violation\\
6. QCD\\
7. Heavy Ion Collisions\\
8. Outlook}


\section{Introduction}

It is not a simple task to summarise in a few pages this exciting Conference that covered all fields in particle
physics. For this reason, may be, the Organisers have scheduled two summary talks \cite{dar}.
 I will try to discuss some of
the results that I found particularly interesting. Even on this limited programme I will
be far from complete. I
apologise to many important contributions (in particular those from the parallel sessions) that I was not
able to take into proper account (although, in order to prepare this talk, I am perhaps the participant who devoted the
shortest time to explore Hanoi, which is a terribly interesting place). I think the Conference has faithfully represented
the situation of particle physics, so that the highlights of the Conference coincide with the important developments in our
field. The main areas where  at present most of the action is are (in an order that goes from the sky more and more down to
earth): astrophysics and cosmology, neutrino oscillations, the search for the Higgs and for new physics beyond the Standard
Model (SM), flavour physics (in particular quark mixings and CP violation), QCD issues and heavy ion physics. I will
discuss the important developments more or less following that order.

\section{Astrophysics and Cosmology}

Experimental progress on the fundamental parameters of the Universe is accumulating at an
impressive rate. In his brilliant introduction to the Conference Peccei \cite{pec} has recalled that the most
basic problems of cosmology are to reconstruct the value of $\Omega$ (the ratio of the density to the critical
density) and of its different components:
\beq
\Omega~=~\Omega_m~+~\Omega_{\Lambda}~=~\Omega_B~+~\Omega_{DM}~+~\Omega_{\Lambda}~^?_=~1 \label{om}
\eeq
and to understand the formation of structures in the Universe and their evolution. As reviewed by Turner
\cite{tur} the progress in recent years in this domain has been really impressive and much more will follow soon. The
list of main achievements starts with the convergence of all evidence to a precise range for
the Hubble constant, a prerequisite for all other progress: $H_0~=~65\pm7$~km/s/Mpc (or, equivalently,
$h=0.65\pm0.07$). If I was asked to nominate the experiments of the year I would probably choose Boomerang and
Maxima: their results support inflation by indicating, from the study of the anisotropies of the cosmic microwave
background and the position of the first acoustic peak, that indeed $\Omega =1$ within an accuracy of about $10\%$.
As the values of
$\Omega _m$ and
$\Omega _{\Lambda}$ are separately extimated by other experiments using diverse techniques, an important
redundancy is obtained. The mass density $\Omega _m$ is measured from the study of the mass distribution within galactic
halos and at larger scales, also using gravitational lensing,  while the cosmological constant term $\Omega _{\Lambda}$ is
extimated from the observation of supernovae at different redshifts. The results are $\Omega _m\sim 0.35$  and
$\Omega _{\Lambda}\sim 0.65$ with a
$20-30\%$ accuracy. The finite value of the cosmological constant component to the energy density is the big news of the
last few years. It implies that the Universe is accelerating. It helps making the determination of the age
of the Universe more consistent among different methods. But it also poses very serious problems: why so small? why now
\cite{mart}?  Most of matter is non baryonic dark matter, as the baryonic  matter which dominates the visible component is
bound from nucleosynthesis \cite{tri} and from the position of the second acoustic peak to be $\Omega _B \lappeq {\rm few}\%$.
Finally another important cosmological parameter is the spectral index that fixes the almost scale invariant behaviour of
mass inhomogeneities at large distances:
$n=0.99\pm0.07$, which is also supporting inflation and the cold dark matter scenario. Of particular relevance for particle
physics is the information on dark matter. Most of it must be cold non baryonic and the best candidates are neutralinos or
axions. With a non zero cosmological constant the need for some hot dark matter component has essentially disappeared
(however, while the $\Lambda CDM$ model is good at large distances, it may face problems at small distances
\cite{far}).  Neutrinos are the most obvious candidates for hot dark matter. From the approximate relation
$\Omega _{\nu} h^2\approx\Sigma m_{\nu}/(100$~eV) and the fact that observed neutrino oscillations imply that at
least one neutrino species has
$m_{\nu}\gappeq 0.05$~eV we have that $\Omega _{\nu}\gappeq 0.001$. On the other side, from the bound $\Omega _{\nu}\lappeq
0.1$, obtained from cosmological observations,  we
obtain that, for three neutrinos with almost equal masses, $m_{\nu} \lappeq 2$~eV.

An interesting problem at the interface between particle physics and astrophysics is that posed by ultra high energy
cosmic rays (UHECR) \cite{lat},\cite{sig}. The main interest for particle physics arises because this effect could
possibly provide a window on new physics. So the first question is whether an astrophysical solution is really
excluded. To my understanding we cannot really be sure that the origin of UHECR cannot be explained in terms of the physics
of such extreme objects like gamma ray bursts or active galactic nuclei. Alternatively in terms of particle physics a not
exotic solution could be formulated in terms of extremely high energy neutrinos producing a Z by interaction with cosmic
background neutrinos. Here the main problems are those of the very large flux of incoming energetic neutrinos (again
astrophysics) and the composition of the predicted spectra. More exciting solutions are in terms of physics near $M_{GUT}$:
very heavy particles with masses near
$M_{GUT}$ (wimpzillas) floating around in the galaxy and occasionally decaying (possibly into $\nu \bar\nu$: see above), producing a UHECR event. Or topological
defects, like cosmic strings, that appear as a particularly promising mechanism. What I think is particularly interesting is
that the predicted spectra in the different models could be tested in the near future.

\section{Neutrino Oscillations}

Neutrino oscillations imply unequal neutrino masses. In analogy to quark mixing, also neutrino mixing arises from
a misalignement of the flavour diagonal and the mass diagonal basis. The smallness of neutrino masses is
explained in terms of the very large scale where lepton number conservation is violated. This is an example of what
Peccei
\cite{pec} called "the very small - very large connection": $m_{small} = m^2/M_{large}$. For Majorana neutrinos,
the light neutrino mass $m_{\nu}$ is related  by the
see-saw formula to the Dirac mass m and the large right-handed neutrino mass M: $m_{\nu} = m^2/M$. Grand
Unified Theories (GUTs) make neutrino masses very plausible. In fact, L and B non conservation is a general feature
of GUTs. The right-handed neutrino $\nu_R$ exist in most GUTs other than minimal SU(5) (for example in the 16 of
SO(10) all fermions of one generation, including $\nu_R$, are exactly accomodated). The large value of $M_{GUT}$ offers
a natural explanation of why neutrino masses are so small.

The evidence for atmospheric neutrino oscillations is rather solid by now\cite{suz}. There are several experiments that
study the effect (SOUDAN2, MACRO, (Super)Kamiokande, K2K...). In particular SK is accumulating a large statistics: sofar
1114 days at about 9 events per day. There are direct signals of oscillations like the up/down asymmetry and the
zenith angular dependence. For example, a test was reported of the L/E dependence of the oscillation function: a
form $\sin^2{(\alpha LE^n)}$ was tested with $\alpha$ and n fitted. The result $n=-1.06\pm0.14$ is obtained. The
transition $\nu_{\mu}\rightarrow\nu_{\tau}$ is found to be the dominant one for atmospheric neutrino oscillations. The
amount of $\nu_{\mu}\rightarrow\nu_e$ is bound by SK, CHOOZ and Palo Verde to $|U_{e3}|\lappeq 0.1$ in the range of
$\Delta m^2$ indicated by SK. In a 2-flavour analysis the transition $\nu_{\mu}\rightarrow\nu_{sterile}$ is excluded
at $99\%$ c.l. (a sterile neutrino is one without weak interactions, so that it is not excluded by LEP). A SK combined
fit gives $\Delta m^2_{atm} = 1.5~-~5~10^{-3}$~eV$^2$ and $\sin^2{\theta_{atm}}\gappeq 0.88$ ($90\%$ c.l.). Note that if
we assume that one neutrino mass (presumably the third generation neutrino mass $m_3$) is much larger than any other,
then 
$\Delta m^2_{atm}=m_3^2$, or $|m_3|\approx 0.05$~eV. If we apply the see-saw formula with $|m_3|\approx 0.05$~eV,
$m\approx 200$~GeV, like the top quark mass or the Higgs vacuum expectation value, we find $M\approx 0.7 ~10^{15}$~GeV, a
value amazingly close to
$M_{GUT}$. This is an impressive confirmation that indeed neutrino masses are related to the scale of lepton number non
conservation. It would be extremely important to confirm the existence of atmospheric neutrino oscillations by a long
baseline laboratory experiment. This is the purpose of the K2K experiment \cite{suz} and there is a great interest on
its results. The analysed events sofar correspond to $2~10^{19}$ protons on target (the goal is $10^{20}$). There is a
promising deficit of events with respect to the non oscillation hypothesis (17 events observed, $29.2\pm^{3.5}_{3.3}$
expected for no oscillations). The rate is perfectly compatible with oscillations with $\Delta m^2_{atm} =
3~10^{-3}$~eV$^2$ and $\sin^2{\theta_{atm}}=1$: in this case the expected rate is $19.3\pm^{2.5}_{2.4}$. But the
statistics is still too low and the significance is only about $2\sigma$.\footnote{At the Osaka Conference, with more
data analysed the numbers became: obs: 29, no oscillations: $40.3\pm^{4.7}_{4.6}$, oscillations with $\Delta m^2_{atm}
= 3~10^{-3}~eV^2$ and $\sin^2{\theta_{atm}}=1$: $26.6\pm^{3.4}_{3.3}$.}. The long baseline investigation of atmospheric
neutrino oscillations will continue by K2K and later by MINOS and OPERA\cite{cam}.

For solar neutrinos \cite{war} the experimental situation and the interpretation in terms of frequency and mixing
angles is less clear. SK reported on new results obtained with a $35\%$ increase in statistics, the threshold energy
lowered down to 5.5~MeV and the background rejection considerably improved. All data were reanalysed and the
systematics reevaluated. The interesting developments can be summarised by saying that the flux deficit is
confirmed ($SK/BP98 = 0.465\pm0.005\pm^{0.015}_{0.013}$), but no other direct signal of oscillations emerges: the
spectrum is compatible with flat ($\chi^2_{flat}/d.o.f.=13.7/17$), the day-night effect is reduced at about
$1.3\sigma$ from zero, and the seasonal variation is compatible with the flux variation from the orbit eccentricity
(i.e. no need of an extra contribution from a modulation of the oscillation rate). Concerning the allowed domains in
the $\Delta m^2_{sun}$, $\sin^2{2\theta _{sun}}$ plane, an optimist may say that we are converging on a smaller and smaller
region. In fact the SMA and the VO solutions are now disfavoured at $95\%$ and what are left are the LMA and LOW
solutions both with large $\sin^2{2\theta  _{sun}}$ and $\Delta m^2_{sun}= 2~10^{-5}~-~10^{-4}$~eV$^2$ (LMA) or $\Delta
m^2_{sun}= ~10^{-7}$~eV$^2$ (LOW). The transition $\nu_e\rightarrow \nu_{sterile}$ is disfavoured at $95\%$, so that the
interpretation is $\nu_e\rightarrow \nu_{\mu},\nu_{\tau}$. A pessimist may fear that soon all solutions will be
excluded, showing that there is an inconsistency either among the experiments or in the theoretical interpretation.
In this respect it is important to keep in mind that the $\Delta m^2$ values of the above solutions are determined by
the experimental result that the observed flux suppression is energy dependent. This is obtained by comparing
experiments with different thresholds. The Cl experiment shows a suppression larger than by a factor of 2, which is
what is found by Ga and water experiments. If the Cl indication was disregarded, then new energy independent solutions
would emerge, with a much more extended range of $\Delta m^2$ and nearly maximal mixing. The experimental situation
will be hopefully clarified with the results from future experiments like SNO, BOREXINO and KamLAND. SNO is already
taking data and the results should come out soon (they have already "seen" the sun through its neutrino radiation).
The ratio NC/CC of the rates of neutral and charged current reactions measured at SNO can provide a direct evidence
of oscillations. Similarly the comparison of the elastic scattering rate at SK with the CC rate in SNO can also
indicate oscillations and by its rate the CC measurement can help distinguishing the different solar neutrino
solutions. BOREXINO can measure in real time the $^7Be$ flux and is sensitive to the VO solution. KamLAND is
sensitive to the LMA solution. Thus it is well possible that the solar neutrino problem will be better clarified in
the near future.

The issue of the additional possible evidence of $\nu_e\rightarrow \nu_{\mu}$ neutrino oscillations from the
LSND experiment is still open \cite{wil}. LSND is neither confirmed nor completely excluded by KARMEN-2. The range of $\Delta
m^2$ implied by the LSND oscillation signal and not excluded by other experiments is at small mixing angle with
$\Delta m^2= 2~10^{-1}~-~2~$eV$^2$ (plus perhaps a small interval near 6~eV$^2$). This range appears to lay much above
the solar and atmospheric neutrino oscillation frequencies, so that the need to accomodate a third frequency would imply the
existence of a fourth light neutrino, that is a sterile neutrino in addition to the established three weakly interacting
flavours. Thus the implication of LSND would be very important. It is clear that the fact that sterile neutrinos are
disfavoured by solar and atmospheric neutrino experiments goes against the LSND claim. The experiment MiniBOONE is designed
to solve this problem starting in '02. There will be a possible follow up by BOONE if the signal is confirmed.

If, pending confirmation, we disregard the LSND evidence, then we can do with the familiar 3 neutrinos and define
$\Delta m^2_{sun}= m^2_2-m^2_1$, $\Delta m^2_{atm}\propto m^2_3-m^2_2$. We have seen that, at present, experiments suggest
that both atmospheric and solar neutrino mixing angles are large or even nearly maximal. There is also evidence that the
electron neutrino does not participate in the atmospheric oscillations, so that the third mixing angle  or, more
precisely, $|U_{e3}|$ is small. These results lead directly to the following approximate form of the
$U_{fi}$ mixing matrix (f=e,$\mu$,$\tau$, i=1,2,3), apart from sign convention redefinitions 
(here we are disregarding CP violation phases: all entries are taken real)
\beq  U_{fi}= 
\left[\matrix{ 1/\sqrt{2}&-1/\sqrt{2}&0 \cr 1/2&1/2&-1/\sqrt{2}\cr
 1/2&1/2&1/\sqrt{2}    } 
\right] ~~~~~.
\label{ufi}
\eeq
In terms of the light neutrino mass matrix, the most likely picture is obtained by assuming that the neutrino mass
eigenvalues are widely different, like for quarks and charged leptons, with dominance of $m_3$: $|m_3| \approx 0.05$~eV$^2>>
|m_{2,1}|$. This hyerarchical situation is well compatible with the see-saw mechanism because the Dirac mass hierarchy tends
to be amplified in this case. This possibility implies that neutrinoless $\beta \beta$ decay is outside the experimental
reach ($m_{ee}$ is proportional to the small masses) and is consistent with baryogenesis through leptogenesis. The
alternative possibility is the case of three almost
perfectly degenerate neutrinos. This is the only choice that could in principle accomodate neutrinos as hot dark matter
with $\Omega _{\nu} \approx 0.1$. According to our previous discussion, the common mass should be
around 1-2~eV. 
A strong constraint arises in the degenerate case from
neutrinoless double beta decay that could be observable in this case. Actually the existing bounds can only be 
satisfied
if solar neutrino oscillations  also occur with nearly maximal mixing. Note that for degenerate masses with $m\sim 1-2$~eV
we need a relative splitting $\Delta m/m\sim
\Delta m^2_{atm}/2m^2\sim 10^{-3}$ to accomodate atmospheric neutrinos and a much smaller one for solar neutrinos. It is
not simple to imagine a natural mechanism compatible with unification and the see-saw mechanism to arrange such a
precise near symmetry.

The prospects for the long range future of neutrino physics were discussed at the Conference \cite{cam}, \cite{gav}. Long
baseline experiments will continue with MINOS in the US and OPERA in Europe. MINOS will study  $\nu_{\mu}$ disappearance
starting from '05. It could observe the oscillation pattern, improve the precision on $\Delta m^2_{23}$ and
$\sin^2{2\theta_{23}}$, be sensitive to $\sin^2{2\theta_{13}}$ and distinguish $\nu_{\mu}\rightarrow \nu_{\tau}$
from $\nu_{\mu}\rightarrow \nu_{sterile}$. The neutrino beam from CERN to the Gran Sasso Laboratory (CNGS) has been
decided in '99. The OPERA experiment has been approved while I was preparing this writeup. Starting the data taking in '05
the main goal of OPERA is to detect $\nu_{\tau}$ appearance (for  $\Delta m^2_{23}\gappeq 10^{-3}$~eV$^2$) and it is also
sensitive to
$\sin^2{2\theta_{23}}$. Other proposed detectors in the Gran Sasso \cite{ric} are ICARUS (now commissioning a 600 t
prototype) and MONOLITH (mainly for studying atmospheric neutrinos, but that could also use the neutrino beam). For the
future the most exciting perspective is given by neutrino factories and very long baseline experiments \cite{gav}.

\section{Higgs and New Physics Searches}

This is certainly the central issue in particle physics today. I start by recalling the context. The SM is in agreement with
all we know from accelerator experiments. Yet nobody believes that the SM is the really fundamental theory. Actually most
theorists think that, even near the weak scale, it must be extended to a more satisfactory theory. All together there is
already a well established standard way beyond the SM: the Minimal Supersymmetric Standard Model (MSSM) near the weak scale,
some form of Grand Unified Theory at $M_{GUT}$ and a really fundamental theory at the Planck scale, $M_P$, including
all gauge interactions and quantum gravity (superstring theory). Of course there are big question marks and alternative
possibilities (examples are compositeness of some sort or, recently, extra dimensions decompactified not far from the weak
scale), but I would say that this is the main benchmark which is proposed to experiment as a reference target. 

The logic
for expecting new physics even close to the weak scale is as follows. The SM does not include quantum gravity which is
certainly important at $M_P$. A simple running of the gauge couplings, assuming no new physics, leads to $M_{GUT}\approx
10^{15}-10^{16}$~GeV not so far from $M_P\approx 10^{19}$~GeV. Thus the SM is at best a low energy effective theory, the
real fundamental theory taking place somewhere between $M_{GUT}$ and $M_P$. But one could conceive the attractive
conjecture that the SM is perhaps valid up to such large energy scales. However, these scales are so large that a low
energy theory must be expecially insensitive to the new physics at the cutoff
$\Lambda$, defined as the scale where the SM is no more valid. A necessary condition is that the low energy theory is
renormalisable. Then the cutoff can be reabsorbed into a redefinition of couplings and masses. But if we see the
renormalisation formalism as a physical requirement of insensitivity to the physics at the cutoff scale, then also the
dependence of masses and couplings on the cutoff must be reasonable. This is not the case for the bosonic sector of the SM.
As well known, the W/Z masses, for example, should be proportional to $\Lambda$ (times a gauge coupling) in the SM and this
poses a clear problem (the hierarchy problem). In comparison fermion masses are stable because their dependence on the
cutoff is as
$m\log{\Lambda}$. Composite models solve the problem by denying that there is a bosonic sector (the Higgs is made of
quarks) at the price of introducing a very strong force that is difficult to hide and that could clash with unification.
Extra dimensions would make the problem less acute by reducing the spread between scales. Supersymmetry solves the problem
because in the symmetric limit the more singular behaviour of bosons is brought down at the level of fermions. 

For broken
supersymmetry the role of the cutoff is played by the splitting of SUSY multiplets. Therefore the smallness of $m_{W/Z}$
requires that spartners are not far from the weak scale and that the SM is invalidated at the TeV scale. It is
interesting that one can construct realisations of the SUSY extended SM that are completely consistent, renormalisable,
perturbatively computable and in agreement with all accelerator experiments like the SM. The MSSM is a simple model of this
kind. In particular the MSSM passes all electroweak precision tests, because for large spartner masses, beyond the present
experimental limits, the radiative corrections in the MSSM are the same within experimental accuracy than those of the SM
with a light Higgs ($m_H \approx 100$~GeV). Since, indeed, all experimental indications ($\sin^2{\theta_{eff}}$, $m_W$, the Z
partial widths...) point towards a light Higgs \cite{rol}, there are at least two fully developed, consistent, and predictive models
that pass all precision SM tests: the SM and the MSSM. But the MSSM has several advantages over the SM not only with
respect to the hierarchy problem, but also as far as the picture of Grand Unification is concerned. It is impressive that
the one-scale SUSY GUT model makes the coupling unification quantitatively exact, which would not be the case in the SM,
given the precisely measured values of the three gauge couplings at the weak scale. Also the non observation of proton decay
is simpler to explain in SUSY GUTS ( although the present Superkamiokande bounds start being very constraining also for these
models). The neutralino is a perfect candidate for cold dark matter (and could be the explanation of the DAMA signal). 

Radiative corrections depend only
logaritmically on $m_H$.  In spite of this small sensitivity, the data are precise enough that one obtains a quantitative
indication of the Higgs mass range from radiative corrections:
\cite{rol} $\log_{10}{m_H({\rm GeV})}=1.88^{+0.28}_{-0.30}$ (or
$m_H=77^{+69}_{-39}$). This result on the Higgs mass is particularly remarkable. The value of
$\log_{10}{m_H({\rm GeV})}$ is right on top of the small window between $\sim 2$ and $\sim 3$ which is allowed by the
direct limit, on the one side, and the theoretical upper limit on the Higgs mass in the minimal SM
$m_H\lappeq 600-800$~GeV, on the other side. We have seen that this result is also perfectly compatible with the MSSM. 
Thus the whole picture of a perturbative theory with a fundamental Higgs is well supported by the data on radiative
corrections. It is important that there is a clear indication for a particularly light Higgs. This is quite encouraging for
the ongoing search for the Higgs particle. If we are interested in other possible forms of new physics other the the SUSY
solution, we can argue in more general terms. If the Higgs couplings are removed from the lagrangian the resulting theory is
non renormalisable. A cutoff
$\Lambda$ must be introduced. In the quantum corrections 
$\log{m_H}$ is then replaced by $\log{\Lambda}$ plus a constant. The precise determination of the associated finite
terms would be lost (that is, the value of the mass in the denominator in the argument of the logarithm). Thus the fact
that, from experiment, one finds $\log{m_H}\sim 2$ is a strong argument in favour of the specific form of the Higgs
mechanism as in the SM or in the MSSM. A heavy Higgs would need some unfortunate conspiracy: the finite terms should
accidentally compensate for the heavy Higgs in the few key parameters of the radiative corrections (mainly $\epsilon_1$ and
$\epsilon_3$). Or additional new physics, for example in the form of effective contact terms added to the minimal SM
lagrangian, should accidentally do the compensation, which again needs some sort of conspiracy \cite{bar}. 

The present limits on the Higgs mass and on the SUSY search as presented by the LEP2 collaborations on July 20 at CERN
were announced here \cite{mop}. For the SM Higgs the present combined limit is $m_H\gappeq 113$~GeV and for the lightest
SUSY Higgs  $m_h\gappeq 90$~GeV (the theoretical upper bound in the MSSM is around 135~GeV) and $\tan{\beta}\gappeq 2.3$.
Also no signals of SUSY particles are found. For example, the neutralino, chargino, s-electron, s-top mass limits are
typically at the values 40~GeV,~103~GeV,~98~GeV,~90~GeV respectively. The s-tau eccess has disappeared. Thus, SUSY behind the
corner has been excluded. Should we start being worried? To some extent yes, as Barbieri explained \cite{bar}. The
MSSM needs some level of fine tuning to explain how comes that we did not see the lightest fringes of the SUSY spectrum. But
while it can certainly be said that LEP could well have discovered the Higgs and some SUSY particles, and in this sense we
are deceived, still, for conceptual reasons, we remain confident that the solution of the Higgs problem and the associated
new physics must be nearby. The hunting will continue at the Tevatron \cite{gra} and the LHC.

\section{Flavour Physics and CP Violation}

The main topics on flavour physics presented at this conference where results on K and B decays of relevance for the
$V_{CKM}$ matrix elements and CP violation plus some presentations on T \cite{ram} and  CPT invariance tests \cite{dos}. The
experiments on K decay were reviewed by Wahl \cite{wal},\cite{arc}. The main experiments involved are CP Lear, KTeV and
NA48. On
$\epsilon '/\epsilon$ the last generation of experiments turned out to be even more difficult than it was expected: the
$\chi^2$ of the available results is still poor. The best values are
$Re(\epsilon '/\epsilon)=(14.0\pm4.3)~10^{-4}$ (NA48) and $Re(\epsilon '/\epsilon)=(28.0\pm4.1)~10^{-4}$ (KTeV) and the
world average is $Re(\epsilon '/\epsilon)=(19.3\pm2.4)~10^{-4}$. The theory is also problematic, because of the
hadronic matrix elements that at the kaon energy scale are completely non perturbative. What appears reasonably
well disintangled is that $\epsilon '/\epsilon\not=0$, so that direct CP violation is established. This is an important
result. The present central value is higher than most theoretical extimates \cite{fab}. But it is not possible to conclude
that the SM is out and new physics is needed, because of the large experimental and theoretical uncertainties. And I
personally doubt that this issue can be completely clarified in the near future. But the potentiality of KTeV and NA48 is
still not exhausted and more data will be collected and/or analysed.  Hopefully $DA\phi NE$ will
eventually send enough luminosity to KLOE for an independent measurement. Thus an improvement in precision is to be
expected. On a related front, progress in the study of rare K decay has been reported \cite{wal},\cite{anz}. In particular NA48 has
presented new results on
$K_S\rightarrow \gamma \gamma$, $K_S\rightarrow \pi^0e^+e^-$ and $K_L\rightarrow e^+e^-e^+e^-$ \cite {anz} and E799 on
$K_L\rightarrow \pi^0e^+e^-$.

On B decays the theoretical introduction was provided by Ball \cite{bal} where the most important channels
for measuring the angles of the CP violation triangle and for the search of new physics in rare decays were presented. In the
SM all CP violating observables must be proportional to J, the Jarlskog invariant (its value is independent of the
parametrization), or equivalently the area of the Bjorken triangle. At present the most direct evidence for
J non vanishing is obtained from the measurement of $\epsilon$ in K decay. Other indirect evidence from all the existing 
measurements of the $V_{CKM}$ parameters is also compatible with this observation (recently important contributions to the
measurement of B decays were obtained at LEP \cite{cas} and at the Tevatron). At B factories the goal is to precisely measure
as many as possible CP violating asymmetries in B decay so as to check that all of them can be reproduced in terms of the
single parameter J as predicted by the SM. The theoretical control of B decays is generally better than for K decays, due to
the larger B mass. But, as recalled by Ball, apart from a few gold plated channels, the extraction of the angles of the
Bjorken triangle is not simple at all and in some cases is model dependent. The most important new developments in this
field are the first results from BaBar
\cite{fer} and Belle \cite{mar}. Both machines are working very well. In their first year of operation, BaBar has
accumulated about 13.5~fb$^{-1}$ of luminosity and Belle 6.2~fb$^{-1}$. BaBar has reported on a measurement of the
branching ratios for the $B->\pi \pi,~K \pi,~KK$ channels. More results are ready for Osaka.

\section{QCD} 

Progress and challenges in perturbative QCD were reviewed by Sterman \cite{ste}. In inclusive processes the QCD
predictions are, in general, in very good agreement with experiment. In less inclusive processes the situation is less
bright and areas of trouble persist. For example, data seem to be in excess of the NLO QCD theory predictions by a factor of
order 2-3 in b production at the Tevatron, as a function of the minimum transverse momentum of the beauty quark \cite{gos}.
A similar excess is also observed at HERA and also in $\gamma-\gamma$ collisions at LEP \cite{gos}. No clear theoretical way
out has been found other than invoking imperfect charm rejection or an exceptionally large impact of higher order terms that
could not sofar be identified and resummed and/or a pile up of many individually small effects. The main advances in
QCD theory are in the technology for computing higher order terms in perturbative expansions. The most crucial stumbling
block is at present the complete analytic calculation of the 3-loop splitting functions of QCD parton evolution. The group
of Vermasereen succeeded in computing the first few moments of singlet and non singlet splitting functions. A complete
calculation of the 3-loop splitting functions appears no more out of the question and probably will be completed in the near future. Another area of
theoretical progress is in the understanding of power corrections and their relation to the large order behaviour of
perturbation theory (renormalons). A different domain is the application of resummation techniques to evaluate the effect of
sequences of particularly large higher order terms in some specific limit. A particularly relevant and interesting case of
this sort is the BFKL resummation of low x contributions in structure functions and their comparison with the HERA data
\cite{bha}. This poses a problem because the corrections are nominally very large but the data are instead in very good
agreement with the NLO QCD fit, showing no apparent signal of these corrections. But the effect of the corrections could be
hidden in the extracted values of the gluon density and of the strong coupling that could be biassed.

\section{Heavy Ion Collisions}

The status of theory \cite{gava} and experiments \cite{gon} in heavy ion collisions have been reviewed at the Conference.
Aspects of QCD at finite density and of colour superconductivity (a phenomenon not relevant for heavy ion collisions but
possibly for neutron stars) were covered also \cite{son},\cite{hie}.  Gonin \cite{gon} has presented the indications collected at the
CERN-SPS for the observation of quark gluon plasma, expecially from charmonium suppression, strangeness enhancement, hadron
yield distributions versus temperature and dilepton excess. I think it is fair to say that there is large amount of
consistency among the different signatures of a phase transition to a new state of matter. Actually the experimental
picture is more definite and sharper than I apriori expected it would be possible to reach. In Pb-Pb collisions at the
CERN-SPS initial energy densities of about
$\epsilon\sim3.5$~GeV/fm$^3$ are obtained.  For thermalised matter this corresponds to temperatures around $T\sim
210-220$~MeV, while the critical temperature for colour deconfinement is extimated around $T\sim170$~MeV (or
$\epsilon\sim0.6~$GeV/fm$^3$). So at CERN one can only explore the region at or just above the phase transition. In a small
range around the critical temperature,
$\epsilon/T^4$ varies by approximately an order of magnitude before flattening to a constant value close to what is expected
for a Stephan-Boltzmann ideal gas of non interacting quarks and gluons. In the case of charmonium suppression, experiments
clearly show that it is visible in Pb-Pb collisions only starting from a definite value of the energy density in the right
range. Actually there is good
quantitative evidence for separate thresholds in
$\epsilon$ for 
$\psi'$,
$\chi$ and $\psi$ suppression which follow each other in the expected order. One in fact expects less-bound states to
dissolve first: $\psi'$,
$\chi$ and $\psi$ should start breaking down at $T_C$, $1.2T_C$ and $1.3 T_C$, respectively, and the observed values are
compatible with this expectation. Enhanced strange particle production in Pb-Pb versus p-Pb collisions offers another
strong indication for the phase transition. In the deconfined phase strange quarks are more easily produced. In
fact the threshold for producing quarks is much lower than that for strange hadrons. Also, the strange quark mass goes
down when chiral symmetry is restored and production of ordinary quarks is suppressed by Fermi statistics. Thus the
observation of large ratios for strange particles in Pb-Pb versus p-Pb or p-Be collisions (the ratios are larger for hadrons
with more strange quarks: about 15 for $\Omega +\bar{\Omega}$) is interpreted as a remnant of the unconfined phase. Moreover
the observed ratios of some 18 hadronic particles measured in Pb-Pb collisions are in agreement with a chemical equilibrium
fireball model with $T\sim 170$~MeV. Finally the dilepton abundance, while well explained by ordinary sources in all other
cases, for Pb-Pb collisions a neat excess is instead observed. In conclusion, putting all indications together, the case for
a new state of matter onsetting at a critical temperature around 170~MeV is reasonably well supported. This evidence makes the
prospects for the new generation of experiments now beginning at RHIC very bright.

\section{Outlook}

Today in particle physics a double approach is followed: from above and from below. From above there are, on the theory
side, quantum gravity (that is superstrings), GUT theories and cosmological scenarios. On the experimental side there
are underground experiments (e.g. searches for neutrino oscillations and proton decay), cosmic ray
observations, space experiments (like COBE, Boomerang, Maxima etc), cosmological observations and so on. From
below, the main objectives of theory and experiment are the search of the Higgs particle, the clarification of the
electroweak symmetry breaking mechanism and the investigation and discovery of physics beyond the Standard Model (typically
supersymmetry, compositeness, large extra dimensions). LEP is about to close and the high energy collider
experimentation will continue at the Tevatron, starting in 2001 and then at the LHC from 2005. Another important
direction of research is aimed at the exploration of the flavour problem: study of CP violation and of rare decays.
The first results from the B factories are the most important new development in this domain. The general expectation
is that new physics must be close by and could be within reach if not for the complexity of the necessary
experimental technology that makes the involved time scale painfully long. 

Finally, as I am the last speaker, I would like, on behalf of all participants, to express our appreciation for the
perfect organisation, the magnificent hospitality and the very pleasant and stimulating atmosphere that we all
enjoyed at this Conference.

I will refer to a talk given at this Conference and to the corresponding written contribution in the Proceedings by simply
listing the name of the speaker.


\begin{thebibliography}{199}



\bibitem{dar} P. Darriulat.
\bibitem{pec} R. Peccei.
\bibitem{tur} M. Turner.
\bibitem{mart} J. Martin.
\bibitem{tri} T. X. Thuan.
\bibitem{far} G. Farrar.
\bibitem{lat} A. Letessier-Selvon.
\bibitem{sig} G. Sigl.
\bibitem{suz} Y. Suzuki.
\bibitem{cam} M. Campanelli.
\bibitem{war} D. Wark.
\bibitem{wil} G. Wilquet.
\bibitem{gav} B. Gavela.
\bibitem{ric} J. Rico.
\bibitem{rol} G. Rolandi.
\bibitem{bar} R. Barbieri.
\bibitem{mop} M. Pepe-Altarelli.
\bibitem{gra} P. Grannis.
\bibitem{ram} N. Ramsey.
\bibitem{dos} M. Doser.
\bibitem{wal} H. Wahl.
\bibitem{arc} R. Arcidiacono.
\bibitem{fab} M. Fabbrichesi.
\bibitem{anz} G. Anzivino.
\bibitem{bal} P. Ball.
\bibitem{cas} C. Caso.
\bibitem{fer} F. Ferroni.
\bibitem{mar} D. Marlow.
\bibitem{ste} G. Sterman.
\bibitem{gos} A. Goshaw.
\bibitem{bha} S. Bhadra.
\bibitem{gava} R. Gavai.
\bibitem{gon} M. Gonin.  
\bibitem{son} D. T. Son.
\bibitem{hie} N. V. Hieu.





\end{thebibliography}
\end{document}